\documentclass[preprint,aps,prb,amsfonts,letterpaper]{revtex4-1}

\usepackage{CJK} 
\usepackage{graphicx, multirow,bm, amsmath, amsfonts}
\usepackage{hyperref}
\usepackage{bbold}
\usepackage[usenames,dvipsnames]{color}

\newcommand{\braket}[3]{\bigl{<}#1\big|#2\big|#3\bigr{>}}
\newcommand{\redbraket}[3]{\bigl{<}#1||#2||#3\bigr{>}}

\begin{document}
\begin{CJK*}{UTF8}{bsmi} 
	\title{Crystal field calculations for transition metal ions by application of an opposing potential}
    \author{Fei Zhou(周非)}
	\email{zhou6@llnl.gov}
    \author{Daniel \r{A}berg}
	\email{aberg2@llnl.gov}
		\affiliation{Physical and Life Sciences Directorate, Lawrence Livermore National Laboratory, Livermore, California 94550, USA}
	\date{\today}

\begin{abstract}
We propose a fully {\em ab initio} method, the opposing crystal potential (OCP), to calculate the crystal field parameters of transition metal impurities in insulator hosts. Through constrained density functional calculations, OCP obtains the constraining Lagrange multipliers, 
 which act as cancellation potential against the crystal field and lead to spherical $d$-electron distribution.
The method is applied to several insulators doped with Mn$^{4+}$ and  Mn$^{2+}$ ions and shown to be in good agreement with experiment.
\end{abstract}
\maketitle
\end{CJK*}

\section{Introduction}

Semiconductor and insulating crystals doped with transition metal ions have found numerous applications in modern technological devices, e.g.\ solid-state laser, solid-state lighting, scintillators and infrared to visible up-conversion. \cite{sugano1970multiplets, avram2013optical} The open $d$ shell of $3d$ dopant ions play a central role in the optical properties of these materials, providing low-lying excited states involving $d^N$ multiplets and tunable optical transitions, which are controlled by the interactions these ions experience in host materials of different chemical compositions and crystal structures. Together with the $d$-electron on-site correlation, these interactions with the crystal  determine the overall energy level schemes and splitting of the electronic states and manifest themselves in the optical absorption and emission spectra. Empirically such interactions are dictated by factors such as covalency of the bonding with ligands as well as geometry of the ligand coordination, i.e.\ number of bonds (coordination number) and the length/angle of bonds.  
The crystal field theory (CFT) \cite{VanVleck1932PR208} is a well established tool for quantifying the ligand environment of localized $d$-electrons.\cite{Wybourne1965Spectroscopic} Beyond transition metal ions in optical materials, CFT finds extensive applications in describing the electronic structure of $f$-electrons of lanthanide and actinide systems, \cite{Newman2000, Liu2005Spectroscopic, Burzo1998RPP1099} including studies on magnetic anisotropy energies for permanent bulk- and single-molecule magnets. \cite{Kuzmin1992PRB8219, Kuzmin1994PRB12533,Rinehart2011CS2078}

The semi-empirical Exchange Charge Model (ECM) \cite{ECM} has been successfully applied to transition metal ions in optical materials to calculate the crystal field parameters (CFP).
First-principles approaches,\cite{Brik-3d-ion, dolg2015computational} including wavefunction-based quantum chemical methods for molecules and clusters and the density functional theory (DFT), have the advantage of predicting CFPs from first-principles without having to fit to experiment.
\cite{Reid2009JAC591, Hu2011JPCM045501, Gaigalas2009LJP403, Novak1994PRB2085, Colarieti-Tosti2002PRB195102} 
However, they are often plagued by the following three technical challenges in the electronic structure of $d/f$-systems. (1) Lack of a fully self-consistent treatment of the $d/f$-charge density. (2) Lack of explicit consideration of strong on-site electronic correlation effects and self-interaction, which may severely impair the accuracy of obtained parameters. Earlier, one of us developed a method to calculate crystal field splitting of lanthanide/actinide oxides \cite{Zhou2011PRB085106, *Zhou2012PRB075124} using self consistent DFT+$U$ \cite{Anisimov1991PRB943} calculations with aspherical self-interaction corrections,\cite{Zhou2009PRB125127} although many calculations of different electronic configurations are required to parametrize the CFT model. Recently Nov\'ak and co-workers \cite{Novak2013PRB205139} developed a DFT-based approach using Wannier wavefunctions  and achieved considerable progress in predicting CFPs of rare earth ions. \cite{Novak2013JPCM446001, Novak2014JMMM228, Novak2014OM414} In their approach the $f$-electrons were treated as core electrons in DFT calculations and correlation effects were considered by introducing an adjustable parameter.
(3) Moreover, compared to $f$-electrons, transition metal $d$-electrons hybridizes considerably more strongly with ligand $p$-electrons, resulting in broader $d$ bands and difficulty identifying the crystal field splitting.

In this paper we propose a fully {\em ab initio} method for the CFPs of transition metal dopants in optical host materials. We constrain the $d$-shell charge density to be spherical by a matrix of Lagrange multipliers that act as a cancellation potential and effectively opposes the DFT crystal field potential to produce a spherical $d$-shell distribution. All CFPs are then solved for using a linear equation of the obtained Lagrange multipliers. The opposing crystal potential (OCP) method is outlined below and  applied to select  semiconductors doped with Mn$^{4+}$ and Mn$^{2+}$ ions. Finally, We will return to address the above questions before make conclusions.

\section{Method} \label{sec:method}
Historically CFPs have been defined prevalently in two different normalization conventions, the so-called Stevens \cite{Stevens1952PPSA209} and Wybourne {\cite{Wybourne1965Spectroscopic}  notations. In this paper we adopt the former one with Steven's operators ${O}_p^k$ and the associated real-valued parameters $B_p^k$. The matrix element $   \langle m  | \hat{H}_{\mathrm{CF}} | m'  \rangle \equiv V_{m m'}$ of the crystal field potential $\hat{H}_{\mathrm{CF}} = \hat{O}_p^k B_p^k$ is:
\begin{eqnarray}
V_{m m'}  &=&  \sum_{p=0, 2, \dots}^{2l} \sum_{k=-p}^{p}  \int \bar{Y}_{l}^{m}  {O}_p^k B_p^k  Y_{l}^{m'} d \Omega \nonumber \\
&\equiv & M_{mm',kp} B_p^k , \ {\mathrm{or}\ } \vec{V} = \mathbb{M} \vec{B}  \label{eq:CF}
\end{eqnarray} 
where $|m \rangle$ designates atomic orbitals (e.g.\ $Y_l^m$ states), and $O_p^k$'s are real-valued spherical harmonics
\begin{align*}
O_2^0 = 4 \sqrt{ \pi/ 5 } Y_2^0,  O_2^{1(-1)} &= -2 \sqrt{{2\pi}/{15}} \Re(\Im) Y_2^1(-1)   , \\
O_2^{2(-2)} &= 4 \sqrt{ {2\pi}/{15}}  \Re(\Im) Y_2^2(-2). 
\end{align*}
The Einstein convention for repeated indices is used through the paper. Alternatively, the Wybourne notation adopt {\it complex} spherical harmonics $C_p^k=\sqrt{4\pi/(2p+1)} Y_p^k$ and {\it complex} CFP $B_p^k$ to represent $\hat{H}_{\mathrm{CF}} = \hat{C}_p^k B_p^k$.

Consider a transition-metal ion in a host material. The total energy of the entire system is
\begin{eqnarray}
E&= \sum_{\sigma, mm'} V_{mm'} n_{mm'}^{\sigma} + E_{\mathrm{ee}}[n] + E_{\mathrm{host}}
\end{eqnarray}
where $n$ is the on-site density matrix of $d$ electrons
\begin{eqnarray}
n_{mm'}^{\sigma} = \sum_{nk} f_{nk}^{\sigma} \langle  \psi_{nk}^{\sigma} | m\rangle \langle m' |\psi_{nk}^{\sigma} \rangle, \label{eq:band2n}
\end{eqnarray}
projected from the Kohn-Sham states $\psi_{nk}^{\sigma}$ with occupancy $f_{nk}^{\sigma}$. The on-site Coulomb interaction $E_{\mathrm{ee}}$ between the strongly correlated $d$ or $f$ electrons represents a formidable challenge for accurate electronic structure calculations. While the Local Density Approximation (LDA) and the Generalized Gradient Approximation (GGA) are well known for their qualitative failure in strongly correlated systems due to lack of consideration of on-site correlation, more advanced methods such as DFT+$U$ \cite{Anisimov1991PRB943} provide explicit treatment of $E_{\mathrm{ee}}$. Finally, $E_{\mathrm{host}}$ is defined as the energy of the host material that does not depend explicitly on the $d$ states. The energy derivative with respect to $n$, or generalized ``chemical potential'' of the $d$-electrons, is
\begin{eqnarray}
\partial E/\partial n_{mm'}^{\sigma} = V_{mm'} + \partial E_{\mathrm{ee}}[n]/\partial n_{mm'}^{\sigma}, \label{eq:chem-pot}
\end{eqnarray}
where the first term represents the desired crystal field potential, and the second term represents on-site contributions.

Our OCP method attempts to calculate $V_{mm'}$ with Eq.~(\ref{eq:chem-pot}) enforcing  a spherical charge distribution $
n^{0 }_{mm'} = \bar{n} \delta_{mm'}
$
 where $\bar n = \mathrm{Tr} n/(4l+2)$ is the average occupancy. Since any physical $E_{\mathrm{ee}}$ is rotationally invariant, the second term in  Eq.~(\ref{eq:chem-pot}) must be spherical
$$
\partial E_{\mathrm{host}}[n]/\partial n_{mm'}^{\sigma} |_{n= \bar{n} \mathbb{1}} = \bar{V}_{\mathrm{ee}} \mathbb{1}. 
$$
Therefore OCP has the advantage of having no explicit dependence on the on-site correlation treatment, as long as a rotationally invariant $E_{\mathrm{ee}}$ is used. To be specific, we propose a constrained, non-spin-polarized DFT calculation,
\begin{eqnarray}
   \min_{\rho} E (\rho)    \ \mathrm{s.t.\ } n[\psi_{nk}[\rho]]  = n_0 \label{eq:minimization} \nonumber
\end{eqnarray}
by introducing Lagrange multipliers $\lambda_{mm'}$ for the constraints and minimizing without constraint
\begin{eqnarray}
F &= E - \lambda_{mm'}(n-n_0)_{mm'} \equiv E + \Delta E^{\mathrm{cs}} \label{eq:Lagrangian}.
\end{eqnarray}
The multiplier matrix $\lambda$ then corresponds to the chemical potential in Eq.~(\ref{eq:chem-pot}) under a spherical $d$-shell and coincides with the $V$ matrix other than a trivial scalar shift that may be absorbed into $B_0^0$:
\begin{eqnarray*}
  \lambda = \partial E/\partial n = V + \bar{V}_{\mathrm{ee}} \mathbb{1}. \label{eq:VV0}
\end{eqnarray*}
The procedures follow the Lagrange multiplier method
\begin{itemize}
\item Determine the average occupancy $\bar n$ by normal DFT. Initialize $n_0= \bar n \mathbb{1}$, step size $\mu$, $\lambda^{(1)}=0$
\item For $k=1, \dots, N$
\begin{enumerate}
      \item Compute matrix $n$ from Eq.~(\ref{eq:band2n}).
      \item At fixed $\lambda^{(k)}$, perform (unconstrained) self-consistent DFT minimization with a non-local constraint potential:
\begin{align}
\Delta \hat{V}^{\mathrm{cs}} = - \lambda_{mm'}  | m \rangle \langle m' |. \label{eq:dV}
\end{align}
The above acts to oppose the potential due to the crystal environment. When converged, it cancels the aspherical components of the crystal potential and yields an evenly occupied $d/f$-shell, i.e. a spherical distribution.
                  \item Update $\lambda^{(k+1)} = \lambda^{(k)} - \mu(n-n_0)$.
      \item If $n-n0$ is sufficiently small, break.
\end{enumerate}
\end{itemize}
Finally, we invert linear equation (\ref{eq:CF}) to obtain CFPs $$\vec B  = \mathbb{M}^{-1} \vec \lambda.$$ Note that for a given $l$ shell, there are $\sum_{p=0, 2, \dots}^{2l} 2p+1 = (l+1)(2l+1) $ unknown $B_p^k$'s (including $B_0^0$), matching exactly $(l+1)(2l+1)$ independent elements of Hermitian matrix $\lambda$, and $\mathbb{M}$ is invertible. Following Nov\'ak, \cite{Novak2013PRB205139}  one may alternatively use the completeness of the spherical harmonic basis set. In particular, any given CFP in the Wybourne normalization can be computed from the Lagrange multiplier matrix $\lambda_{mm'}$ according to
\begin{align}
B^k_q = \frac{(-1)^{k+q}}{2k+1}{\redbraket{\ell}{C^k}{\ell}^2}\sum_{mm'} \lambda_{mm'}\braket{\ell m'}{C^k_q}{\ell m }.
\end{align}

In this work, we use experimental values for the cell metric. One transition metal ion is embedded in a supercell of about 70-120 atoms.
We use the Perdew-Becke-Ernzerhof (PBE)  \cite{Perdew1996PRL3865} parametrization of GGA for the OCP calculations, projector augmented-wave (PAW) potentials,\cite{Blochl1994PRB17953} 
and no symmetry constraints 
as implemented in the {\sc VASP} package. \cite{Kresse1999PRB1758} Brillouin zone integration was carried out on $k$-point meshes of at least $2\times 2 \times 2$ with Gaussian broadening of width 0.05 eV. Unless otherwise specified, all internal coordinates of the supercells were relaxed using GGA+$U$ to better treat possible Jahn-Teller structural distortions. We chose atomic basis functions $|m \rangle$ represented by normalized, real-valued spherical harmonics \cite{Zhou2009PRB125127} and ignored spin-orbit coupling so that all matrices $V$, $\lambda$ and $\mathbb M$ are real.

\section{Results } \label{sec:results}
First we study octahedrally coordinated Mn$^{4+}$ used in red phosphors, in which Mn$^{4+}$ ($3d^3$) has a $^4 A _{2g}$ ground state configuration. Depending on the competition between crystal field and Coulomb repulsion, the first excited state is either $^3 T_2$ or $^2 E_g$. The red luminescence is ascribed to the $^2E_g \rightarrow ^4A_{2g} + h \nu$ emission, which is mainly controlled by the free-ion parameters and insensitive of CFPs. Given the high effective positive charge and small ionic radius of Mn$^{4+}$, large CFPs are expected.

Fig.~\ref{fig:SrTiO3-Mn} shows the convergence of our method applied to perovskite SrTiO$_3$:Mn, a rare-earth-free phosphor material. \cite{Bryknar2000JL605} Given the $m\bar{3}m$ point group of the substitutional tetravalent site (Ti$^{4+}$), a crystal field of cubic symmetry for the Mn$^{4+}$ ion is expected. The multipliers $\lambda$ indeed become triply degenerate $\lambda(t_{2g})$  and doubly-degenerate $\lambda(e_{g})$, while the occupancies start with $n(t_{2g}) > n(e_{g})$ and gradually converge towards a uniform $\bar n$, which was set to 0.477 according to initial structurally relaxed GGA+$U$ calculations. Upon convergence, the added potential $\Delta \hat{V}^{\mathrm{cs}}$ in Eq.~\ref{eq:dV}} reaches a splitting $\lambda(e_{g}) -\lambda n(t_{2g})=2.386$ eV and effectively cancels out the crystal field potential on the transition metal ion, leading to overall equal occupancy. Comparing results using step sizes $\mu$=3 eV (solid curves) and 2 eV (dashed), $\mu$ is found to affect only the convergence speed, not the final results.
\begin{figure}[htbp] 
	\includegraphics[width=0.9 \linewidth]{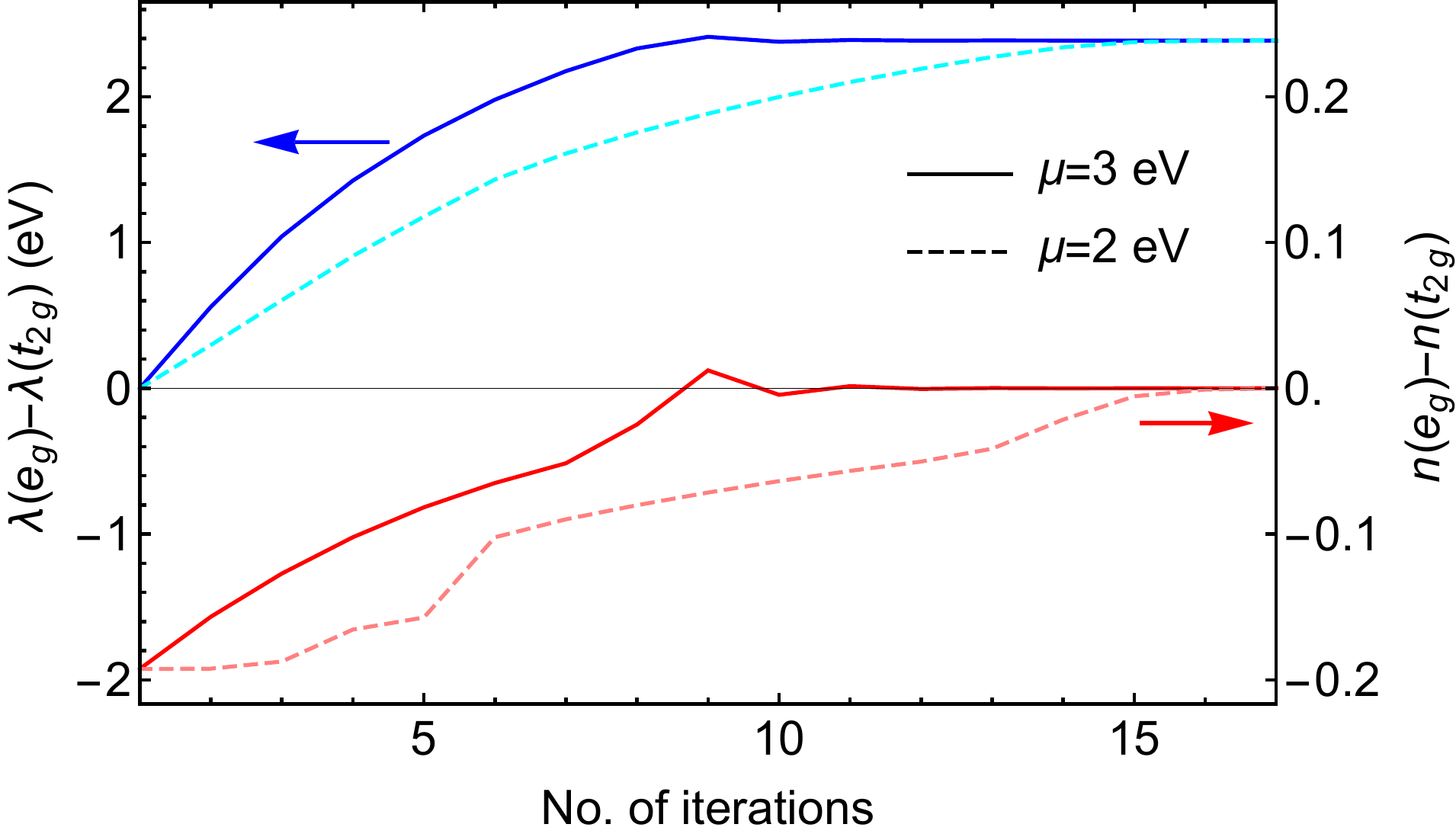} 
	\caption{For SrTiO$_3$:Mn, the $t_{2g}-e_g$ difference in the multipliers (upper blue curves)  and occupancies (lower red), respectively.
				} \label{fig:SrTiO3-Mn} 
\end{figure}

\begin{figure}[htbp] 
	\includegraphics[width=0.98 \linewidth]{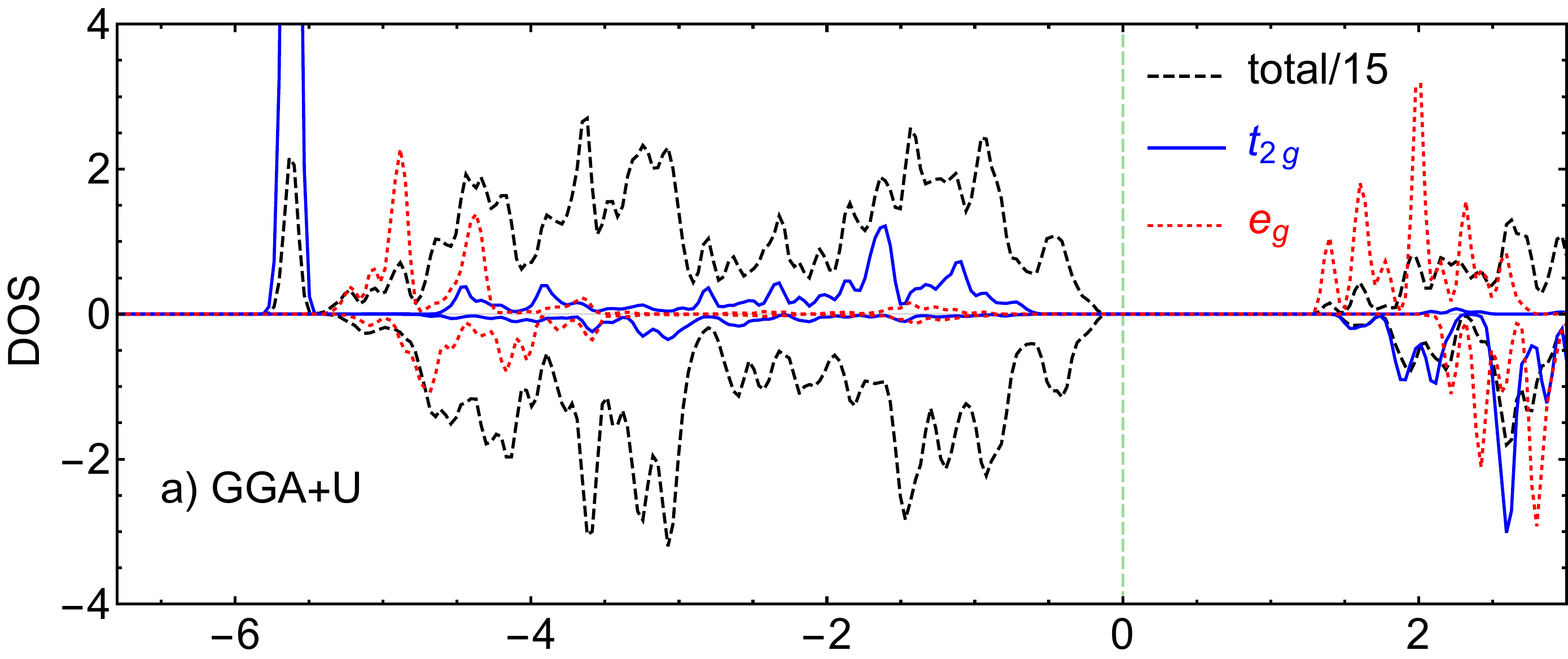} 
	\includegraphics[width=0.98 \linewidth]{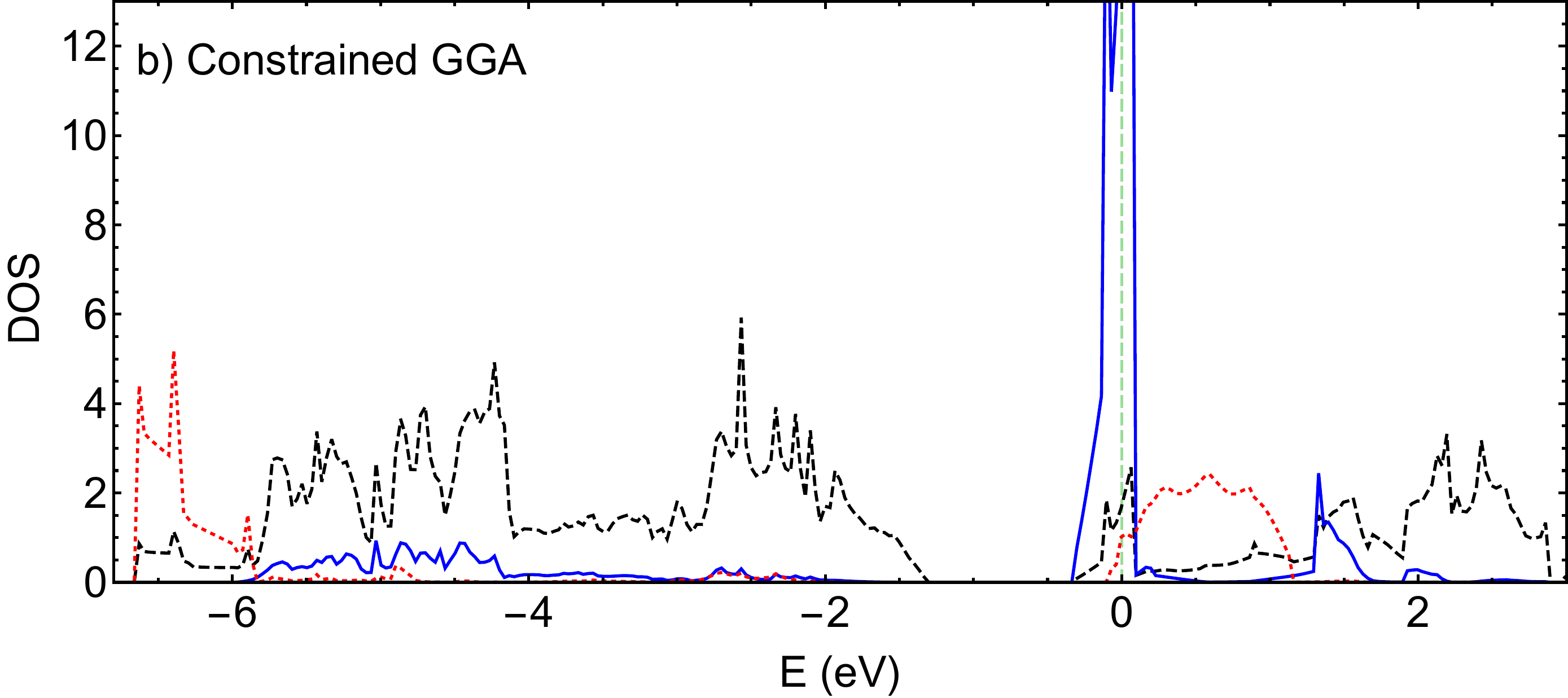} 
	\caption{DOS of SrTiO$_3$:Mn. (a) Normal spin-polarized GGA+$U$ and (b) Our OCP method using constrained GGA calculations. The Fermi level is set to zero. } \label{fig:SrTiO3-Mn-dos} 
\end{figure}

To shed light on the effects of our approach on the electronic structure of SrTiO$_3$:Mn, Fig.~\ref{fig:SrTiO3-Mn-dos}a shows the total and Mn $t_{2g}$/$e_g$ decomposed density of states (DOS) from a normal spin-polarized GGA+$U$ calculation ($U$=5 eV, $J$=1 eV). The strong on-site Coulomb interactions give rise to a charge-transfer band gap and sizable exchange splitting with the Mn$^{4+}$ ion in the $ t_{2g}^3$ majority high-spin configuration. The $p$-$d$ hybridization leads to broad bands with $d$ characters,  broader for $e_g$ with stronger hybridization than $t_{2g}$. In the minority spin channel, both empty $t_{2g}$ and $e_g$ bands are so wide due to hybridization that they lie in approximate the same energy range above the Fermi level, in contradiction to the conventional crystal field picture of clear-cut $t_{2g}$-$e_g$ splitting. Therefore it is infeasible to assign specific energies to these crystal field levels and extract CFPs from Kohn-Sham eigen-energies.  

Fig.~\ref{fig:SrTiO3-Mn-dos}b shows the results from constrained  non-spin polarized PBE calculations, where all five $3d$ orbitals, including $t_{2g}$ and $e_g$, become equally occupied as the crystal potential is effectively canceled by the applied potential $-\lambda$ in Eq.~\ref{eq:dV}. Note that the five $3d$ states are not degenerate but spread out, since in a crystal they contribute to $p$-$d$ bands, not five flat levels. For example, the bonding $p$-$e_g$ bands are well below the Fermi level ($<-6$ eV) whereas the anti-bonding $p$-$e_g$ bands are near the Fermi level.  Our method makes use of well-defined occupancies, not ill-defined CF energy levels. The constrained calculation in Fig.~\ref{fig:SrTiO3-Mn-dos}b leads to metallic electronic structure as an artifact of the GGA functional used.

Table~\ref{tab:CF-vs-xc} shows the $t_{2g}$-$e_g$ splitting 10$Dq$ of SrTiO$_3$:Mn calculated with LDA and GGA. In all cases $10Dq$ increases when one allows for ionic relaxation using GGA+$U$ with decreased Mn-O bond length of 1.906 \AA\ from 1.949 \AA\ in the unperturbed host.  The PBE prediction of 10$Dq$ of the relaxed structure is about 6\% larger than the value 2.25 eV fitted to experimental spectra. The LDA results are almost the same as PBE.  A direct comparison of theory to experimental $d^3$ energy levels requires free-ion parameters (e.g.\ Racah parameters) in addition to CFPs and is not attempted in this work. 
 \begin{table}[htbp] 
	\begin{ruledtabular}
		\begin{tabular}{|c|ccc|} 
		& LDA	& PBE	& Expt.	 \\ 
		\hline
no relax	&  2.181  	& 2.168  	&   \\
relaxed		&   2.398 	& 2.386  	&  2.25 \cite{Bryknar2000JL605} \\
		\end{tabular}
	\end{ruledtabular}
	\caption{Calculated 10$Dq$ in eV for SrTiO$_3$:Mn before and after ionic relaxation.
	\label{tab:CF-vs-xc}} 
\end{table}

\begin{figure}[tbph]
\includegraphics[width=0.78 \linewidth]{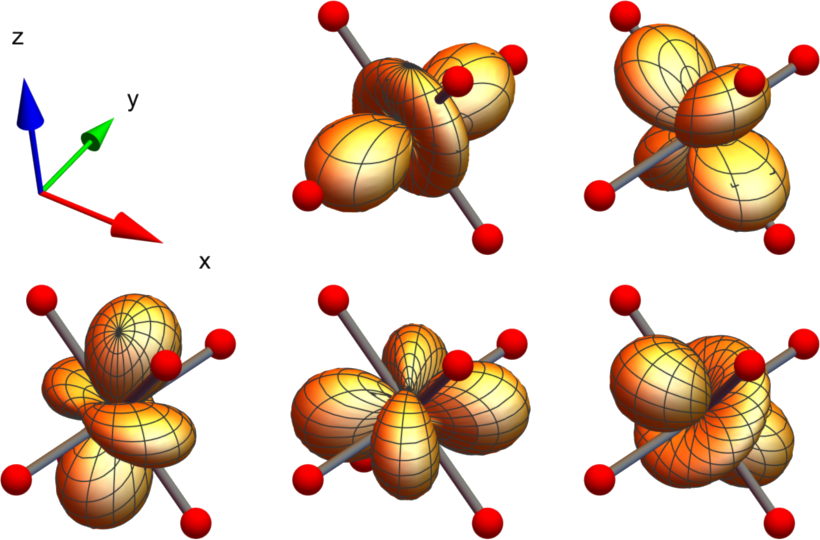}
\caption{The $d^1$ eigenstates according to CFPs of MnO$_6$ in CaZrO$_3$. The upper/lower states are (approximately) $e_g$/$t_{2g}$.}
\label{fig:CaZrO3-d1CF}
\end{figure}

Further tests on several insulators with Mn$^{4+}$ substitution on tetravalent cation sites are shown in Table \ref{tab:Mn4-CF}, including both oxides and fluorides. For non-cubic materials, the 4th eigenvalue of the $\lambda_{mm'}$ matrix is given as 10$Dq$. In Na$_2$SiF$_6$ we list calculated CFPs for both kinds of Si sites: the $1a$ sites with point group $D_3$, and $2d$ sites with $C_3$. In the relaxed configurations, Mn$^{4+}$ is more stable on the $C_3$ sites by 0.06 eV than on the former. Note again that the literature values of $Dq$ are fitted to experiment, not direct experimental observations. Overall good agreement with experimental fitted values is obtained. Our approach reproduces the trend in the crystal field strength very well, with a slight systematic overestimation. The octahedral Zr$^{4+}$ site in CaZrO$_3$ has a low $\bar 1$ symmetry, resulting in a large number of CFPs. The values in Table \ref{tab:Mn4-CF} for CaZrO$_3$:Mn are given directly in the relaxed supercell without re-orientation of axes and are validated by the fact that the corresponding $d^1$ crystal field levels shown in Fig.~\ref{fig:CaZrO3-d1CF}, i.e.\ eigenstates of the corresponding $\lambda$ matrix, are properly aligned with the rotated low-symmetry MnO$_6$ octahedron. For example, the two higher $d^1$ eigenstates have the familiar shape of $d_{z^2}$ and $d_{x^2-y^2}$ with their lobes against the ligand oxygen atoms.
\begin{table}[htbp] 
	\begin{ruledtabular}
		\begin{tabular}{|c|ccccc|} 
		& SrTiO$_3$	&  CaZrO$_3$ & K$_2$SiF$_6$	& \multicolumn{2}{c|}{Na$_2$SiF$_6$}	  \\ 
		\hline
Sp.\ Grp.		&$Pm\bar{3}m$		&$Pcmn$	&	$ F m \bar{3} m$	& \multicolumn{2}{c|}{$P321$} \\
\cline{5-6}
Pt.\ Grp.		&$m\bar{3}m$		&$\bar 1$	&	$ m \bar{3} m$	& $32.$ (1$a$)	& $3..$ (2$d$) \\ \hline
$B_2^{-2}$		&    	&-1674  		&   	&       	& \\
$B_2^{-1}$		&    	&-214   	&       &	& \\
$B_2^{0}$		&    	&729   	&       &-433	& 51 \\
$B_2^{1}$		&    	&-629   	&       &	& \\
$B_2^{2}$		&    	&1959  		&   	&       	& \\
$B_4^{-4}$		& 		&-15496  	& &       	& \\
$B_4^{-3}$		& 		&2400  	& &       	& \\
$B_4^{-2}$		& 		&-11981  	& &       	& \\
$B_4^{-1}$		& 		&-8652  	& &       	& \\
$B_4^{0}$		&5053 	&-929  	&6187 &  -4118     	&-4116 \\
$B_4^{1}$		& 		&17241  	& &       	& \\
$B_4^{2}$		&    	&19336  	&   	&       	& \\
$B_4^{3}$		& 		&-24507  	&	 & -11923     	&-118270 \\
$B_4^{4}$		&25263  &11190  	&30932   	&       	& \\ \hline
$Dq$		&  1925 	&1869 	&  2357 	& 2404      	&2387 \\
Prev.		& 1818    	& 1850  	& 2323   	&   \multicolumn{2}{c|}{1970; 2193} \\
Ref.		& \onlinecite{Bryknar2000JL605}	& \onlinecite{Brik2013EJSSSTR148}  	& \onlinecite{Brik2013JL69}  	&  \multicolumn{2}{c|}{\onlinecite{Xu2009JAP013525}; \onlinecite{Brik2013JL69} 
	\footnote{According to the CFPs in Ref.~\onlinecite{Brik2013JL69}}}
		\end{tabular} \\
	\end{ruledtabular}
	\caption{Calculated CFPs and splitting in cm$^{-1}$ for octahedrally coordinated Mn$^{4+}$ compared to literature values fitted to experimental measurements. The symmetry of substitutional sites is also shown.
	\label{tab:Mn4-CF}} 
\end{table}

Compared to its tetravalent counterpart, the Mn$^{2+}$ ion is characterized by a wide range ($\sim$500--700 nm) of luminescence color, tunable by the crystal field and crystal structure (green in tetrahedral sites, orange/red in octahedral sites). \cite{Curie1974JCP3048, Palumbo1970JES1184, Palumbo1971JES1159} Table \ref{tab:Mn2-CF} shows the results for two chalcogenides (oxide and sulfide) and two halides (fluoride and chloride). Due to the smaller effective charge and larger ionic radius than tetravalent Mn, the overall splitting $Dq$ is considerably smaller.    We again found good agreement with experiment with small systematic overestimation of $Dq$ except CaCl$_2$, for which our prediction of $Dq = 702.5$ cm$^{-1}$ is significantly larger than the only experimental data \cite{Castaneda2005OM1456} but agrees better with $Dq = 718$ cm$^{-1}$ in MnCl$_2$ and the general trend for the crystal field of octahedrally coordianted Mn$^{2+}$ observed in Ref.~\onlinecite{Curie1974JCP3048}. For ZnS our prediction is in better agreement with Ref.~\onlinecite{Whiffen2014JL133} than an earlier fitting in Ref.~\onlinecite{Chen2001JAP1120}.
\begin{table}[htbp] 
	\begin{ruledtabular}
\begin{tabular}{|c|cccc|} 
		& MgGa$_2$O$_4$ & ZnS	& ZnF$_2$	 & CaCl$_2$ 	 \\ 
		\hline
Sp.\ Grp.		&$Fd\bar 3m$ & $F \bar{4}3m$		&$P4_2/mnm$ 			& $P n n m$	\\ 
Pt.\ Grp.		&$\bar{4}3m$		&$\bar{4}3m$	&	$ m m m$	& $..2/m$	\\
Coord.	&	tet &tet		&oct		& oct	\\	\hline		
$B_2^{-2}$		&    	&   	& -1654 	&  -96 	\\
$B_2^{0}$		&    	&   	& 229 	&  247 	\\
$B_2^{2}$		&    	&   	&  	&  122 	\\
$B_4^{-4}$		&    	&   	&  	&  2183 	\\
$B_4^{-2}$		&    	&   	& 13880 	&  9444 	\\
$B_4^{0}$	& -1522	& -1487     	& -499  	 	&-522  \\
$B_4^{2}$		&    	&   	&  	&  1864 	\\
$B_4^{4}$	& -7611	& -7436    	& -8313  	  	&  -5293 	\\ \hline
$Dq$	& 580 	& 567   	& 1020 		&  703 	\\
Prev.		& 520 	& 667; 502  	&  930 	&  507.3      	\\
Ref.	& \onlinecite{Palumbo1970JES1184} 	& \onlinecite{Chen2001JAP1120}; \onlinecite{Whiffen2014JL133}   	&  \onlinecite{Palumbo1971JES1159} 		&   \onlinecite{Castaneda2005OM1456}	\\
\end{tabular}
		\end{ruledtabular}
	\caption{The same as Table~\ref{tab:Mn4-CF} for Mn$^{2+}$ ion. The coordination is either tetrahedral or octahedral.
	\label{tab:Mn2-CF}} 
\end{table}

\section{Summary} \label{sec:conclusion}
Let's revisit the challenges of CFP calculations with DFT outlined earlier and summarize our treatment:
\begin{enumerate}
\item Charge self-consistency. Our OCP method is carried out fully self-consistently without freezing $d/f$-electrons into the core.  The obtained CFPs therefore directly correspond to the crystal environment that the ion experience given a spherical density matrix $\bar{n} \mathbb{1}$. Since the main purpose of the crystal field model is to describe various excited states with completely different spin and charge distribution, our choice of a spherical, non-spin polarized reference distribution is justified as an {\it unbiased average} of $d^N$ configurations and validated by excellent agreement with experiment.
\item On-site correlation effects. Our approach does not explicitly depend on and to a large extent circumvents the complication of on-site correlation for $d$ or $f$-electrons, since any physical exchange-correlation yields a uniform potential shift in our method that gets absorbed into the spherical part ($B_0^0$) of the crystal field. We were able to use local approximations (LDA, GGA) in this work, despite the expected deficiency of GGA for strongly correlated materials. Use of more advanced methods may potentially mildly improve the results.
\item Primarily for $d$-electrons, the difficulty in identifying $d$ bands due to strong $d$-$p$ hybridization. Our method relies on the projected on-site occupancy, which are always well-defined, and does not suffer from the ill-defined assignment of broad energy bands to local $d$-states. 
\end{enumerate}

Our method for CFPs assumes a spherical $d/f$-electron charge distribution, which remains reasonable as long as the charge distribution of the host marial is not significantly altered. For $3d$ ions, the $d$-electrons are relatively localized and we expect this assumption to be valid. In systems with strong $d$-$p$ hybridization and more delocalized $d$-electrons, this assumption needs to be reassessed. However, in those cases the whole notion of the crystal field models is questionable and may have to be replaced by more elaborate theory anyway.

In summary, we developed a parameter-free {\it ab initio} method, the opposing crystal potential, for crystal field calculation based on constrained DFT, with crystal field parameters extracted from the constraining Lagrange multipliers, which effectively oppose the crystal field potential and yield spherical $d/f$-electron distribution. Good agreement with experiment was demonstrated on transition metal ions Mn$^{4+}$ and Mn$^{2+}$ used in rare-earth free phosphors. The approach is highly efficient with approximate the same computational cost as conventional DFT calculations times a small pre-factor towards convergence of the Lagrange multipliers. It is equally applicable to crystals of low or high symmetry and can be readily implemented in general-purpose DFT codes that support on-site non-local potentials (e.g.\ DFT+$U$). Results on rare earth ions will be published separately.

\begin{acknowledgments}
This work was performed under the auspices of the U.S. Department of Energy by Lawrence Livermore National Laboratory under Contract No. DE- AC52-07NA27344. This work is supported by the Critical Materials Institute, an Energy Innovation Hub funded by the U.S. Department of Energy, Office of Energy Efficiency and Renewable Energy, Advanced Manufacturing Office. We thank Babak Sadigh for helpful discussions. \cite{}
\end{acknowledgments}


\begin{thebibliography}{40}\makeatletter
\providecommand \@ifxundefined [1]{ \@ifx{#1\undefined}
}\providecommand \@ifnum [1]{ \ifnum #1\expandafter \@firstoftwo
 \else \expandafter \@secondoftwo
 \fi
}\providecommand \@ifx [1]{ \ifx #1\expandafter \@firstoftwo
 \else \expandafter \@secondoftwo
 \fi
}\providecommand \natexlab [1]{#1}\providecommand \enquote  [1]{``#1''}\providecommand \bibnamefont  [1]{#1}\providecommand \bibfnamefont [1]{#1}\providecommand \citenamefont [1]{#1}\providecommand \href@noop [0]{\@secondoftwo}\providecommand \href [0]{\begingroup \@sanitize@url \@href}\providecommand \@href[1]{\@@startlink{#1}\@@href}\providecommand \@@href[1]{\endgroup#1\@@endlink}\providecommand \@sanitize@url [0]{\catcode `\\12\catcode `\$12\catcode
  `\&12\catcode `\#12\catcode `\^12\catcode `\_12\catcode `\%12\relax}\providecommand \@@startlink[1]{}\providecommand \@@endlink[0]{}\providecommand \url  [0]{\begingroup\@sanitize@url \@url }\providecommand \@url [1]{\endgroup\@href {#1}{\urlprefix }}\providecommand \urlprefix  [0]{URL }\providecommand \Eprint [0]{\href }\providecommand \doibase [0]{http://dx.doi.org/}\providecommand \selectlanguage [0]{\@gobble}\providecommand \bibinfo  [0]{\@secondoftwo}\providecommand \bibfield  [0]{\@secondoftwo}\providecommand \translation [1]{[#1]}\providecommand \BibitemOpen [0]{}\providecommand \bibitemStop [0]{}\providecommand \bibitemNoStop [0]{.\EOS\space}\providecommand \EOS [0]{\spacefactor3000\relax}\providecommand \BibitemShut  [1]{\csname bibitem#1\endcsname}\let\auto@bib@innerbib\@empty
\bibitem [{\citenamefont {Satoru~Sugano}(1970)}]{sugano1970multiplets}  \BibitemOpen
  \bibfield  {author} {\bibinfo {author} {\bibfnamefont {H.~K.}\ \bibnamefont
  {Satoru~Sugano}, \bibfnamefont {Yukito~Tanabe}},\ }\href
  {https://books.google.com/books?id=8SbsjFx1MbwC} {\emph {\bibinfo {title}
  {Multiplets of Transition-Metal Ions in Crystals}}}\ (\bibinfo  {publisher}
  {Academic},\ \bibinfo {address} {New York},\ \bibinfo {year}
  {1970})\BibitemShut {NoStop}\bibitem [{\citenamefont {Avram}\ and\ \citenamefont
  {Brik}(2013)}]{avram2013optical}  \BibitemOpen
  \bibinfo {editor} {\bibfnamefont {N.~M.}\ \bibnamefont {Avram}}\ and\
  \bibinfo {editor} {\bibfnamefont {M.~G.}\ \bibnamefont {Brik}},\ eds.,\ \href
  {\doibase 10.1007/978-3-642-30838-3} {\emph {\bibinfo {title} {Optical
  Properties of 3d-Ions in Crystals: Spectroscopy and Crystal Field
  Analysis}}}\ (\bibinfo  {publisher} {Springer},\ \bibinfo {address}
  {Heidelberg},\ \bibinfo {year} {2013})\BibitemShut {NoStop}\bibitem [{\citenamefont {Van~Vleck}(1932)}]{VanVleck1932PR208}  \BibitemOpen
  \bibfield  {author} {\bibinfo {author} {\bibfnamefont {J.~H.}\ \bibnamefont
  {Van~Vleck}},\ }\href@noop {} {\bibfield  {journal} {\bibinfo  {journal}
  {Phys. Rev.}\ }\textbf {\bibinfo {volume} {41}},\ \bibinfo {pages} {208}
  (\bibinfo {year} {1932})}\BibitemShut {NoStop}\bibitem [{\citenamefont {Wybourne}(1965)}]{Wybourne1965Spectroscopic}  \BibitemOpen
  \bibfield  {author} {\bibinfo {author} {\bibfnamefont {B.~G.}\ \bibnamefont
  {Wybourne}},\ }\href@noop {} {\emph {\bibinfo {title} {Spectroscopic
  Properties of Rare Earths}}}\ (\bibinfo  {publisher} {Interscience},\
  \bibinfo {address} {New York},\ \bibinfo {year} {1965})\BibitemShut {NoStop}\bibitem [{\citenamefont {Newman}\ and\ \citenamefont {Ng}(2000)}]{Newman2000}  \BibitemOpen
  \bibfield  {author} {\bibinfo {author} {\bibfnamefont {D.}~\bibnamefont
  {Newman}}\ and\ \bibinfo {author} {\bibfnamefont {B.}~\bibnamefont {Ng}},\
  }\href@noop {} {\emph {\bibinfo {title} {Crystal Field Handbook}}}\ (\bibinfo
   {publisher} {Cambridge University Press},\ \bibinfo {address} {Cambridge},\
  \bibinfo {year} {2000})\BibitemShut {NoStop}\bibitem [{\citenamefont {Liu}\ and\ \citenamefont
  {Jacquier}(2005)}]{Liu2005Spectroscopic}  \BibitemOpen
  \bibinfo {editor} {\bibfnamefont {G.}~\bibnamefont {Liu}}\ and\ \bibinfo
  {editor} {\bibfnamefont {B.}~\bibnamefont {Jacquier}},\ eds.,\ \href@noop {}
  {\emph {\bibinfo {title} {Spectroscopic properties of rare earths in optical
  materials}}}\ (\bibinfo  {publisher} {Springer},\ \bibinfo {address} {New
  York},\ \bibinfo {year} {2005})\BibitemShut {NoStop}\bibitem [{\citenamefont {Burzo}(1998)}]{Burzo1998RPP1099}  \BibitemOpen
  \bibfield  {author} {\bibinfo {author} {\bibfnamefont {E.}~\bibnamefont
  {Burzo}},\ }\href@noop {} {\bibfield  {journal} {\bibinfo  {journal} {Rep.
  Prog. Phys.}\ }\textbf {\bibinfo {volume} {61}},\ \bibinfo {pages} {1099}
  (\bibinfo {year} {1998})}\BibitemShut {NoStop}\bibitem [{\citenamefont {Kuz'min}(1992)}]{Kuzmin1992PRB8219}  \BibitemOpen
  \bibfield  {author} {\bibinfo {author} {\bibfnamefont {M.~D.}\ \bibnamefont
  {Kuz'min}},\ }\href@noop {} {\bibfield  {journal} {\bibinfo  {journal} {Phys.
  Rev. B}\ }\textbf {\bibinfo {volume} {46}},\ \bibinfo {pages} {8219}
  (\bibinfo {year} {1992})}\BibitemShut {NoStop}\bibitem [{\citenamefont {Kuz'min}\ and\ \citenamefont
  {Coey}(1994)}]{Kuzmin1994PRB12533}  \BibitemOpen
  \bibfield  {author} {\bibinfo {author} {\bibfnamefont {M.~D.}\ \bibnamefont
  {Kuz'min}}\ and\ \bibinfo {author} {\bibfnamefont {J.~M.~D.}\ \bibnamefont
  {Coey}},\ }\href@noop {} {\bibfield  {journal} {\bibinfo  {journal} {Phys.
  Rev. B}\ }\textbf {\bibinfo {volume} {50}},\ \bibinfo {pages} {12533}
  (\bibinfo {year} {1994})}\BibitemShut {NoStop}\bibitem [{\citenamefont {Rinehart}\ and\ \citenamefont
  {Long}(2011)}]{Rinehart2011CS2078}  \BibitemOpen
  \bibfield  {author} {\bibinfo {author} {\bibfnamefont {J.~D.}\ \bibnamefont
  {Rinehart}}\ and\ \bibinfo {author} {\bibfnamefont {J.~R.}\ \bibnamefont
  {Long}},\ }\href@noop {} {\bibfield  {journal} {\bibinfo  {journal} {Chemical
  Science}\ }\textbf {\bibinfo {volume} {2}},\ \bibinfo {pages} {2078}
  (\bibinfo {year} {2011})}\BibitemShut {NoStop}\bibitem [{\citenamefont {Malkin}(1987)}]{ECM}  \BibitemOpen
  \bibfield  {author} {\bibinfo {author} {\bibfnamefont {B.~Z.}\ \bibnamefont
  {Malkin}},\ }\enquote {\bibinfo {title} {Crystal field and electron-phonon
  interaction in rare-earth ionic paramagnets},}\ in\ \href@noop {} {\emph
  {\bibinfo {booktitle} {Spectroscopy of solids containing rare earth ions}}},\
  \bibinfo {editor} {edited by\ \bibinfo {editor} {\bibfnamefont {R.~M.~M.}\
  \bibnamefont {A.~A.~Kaplyanskii}}}\ (\bibinfo  {publisher} {North-Holland},\
  \bibinfo {address} {Amsterdam},\ \bibinfo {year} {1987})\ pp.\ \bibinfo
  {pages} {33--50}\BibitemShut {NoStop}\bibitem [{\citenamefont {Brik}(2013)}]{Brik-3d-ion}  \BibitemOpen
  \bibfield  {author} {\bibinfo {author} {\bibfnamefont {M.}~\bibnamefont
  {Brik}},\ }in\ \href {\doibase 10.1007/978-3-642-30838-3_6} {\emph {\bibinfo
  {booktitle} {Optical Properties of 3d-Ions in Crystals: Spectroscopy and
  Crystal Field Analysis}}},\ \bibinfo {editor} {edited by\ \bibinfo {editor}
  {\bibfnamefont {N.~M.}\ \bibnamefont {Avram}}\ and\ \bibinfo {editor}
  {\bibfnamefont {M.~G.}\ \bibnamefont {Brik}}}\ (\bibinfo  {publisher}
  {Springer},\ \bibinfo {address} {Heidelberg},\ \bibinfo {year} {2013})\ pp.\
  \bibinfo {pages} {203--250}\BibitemShut {NoStop}\bibitem [{\citenamefont {Dolg}(2015)}]{dolg2015computational}  \BibitemOpen
  \bibinfo {editor} {\bibfnamefont {M.}~\bibnamefont {Dolg}},\ ed.,\ \href
  {https://books.google.com/books?id=K2y5BgAAQBAJ} {\emph {\bibinfo {title}
  {Computational Methods in Lanthanide and Actinide Chemistry}}}\ (\bibinfo
  {publisher} {Wiley},\ \bibinfo {address} {New York},\ \bibinfo {year}
  {2015})\BibitemShut {NoStop}\bibitem [{\citenamefont {Reid}\ \emph {et~al.}(2009)\citenamefont {Reid},
  \citenamefont {Duan},\ and\ \citenamefont {Zhou}}]{Reid2009JAC591}  \BibitemOpen
  \bibfield  {author} {\bibinfo {author} {\bibfnamefont {M.~F.}\ \bibnamefont
  {Reid}}, \bibinfo {author} {\bibfnamefont {C.-K.}\ \bibnamefont {Duan}}, \
  and\ \bibinfo {author} {\bibfnamefont {H.}~\bibnamefont {Zhou}},\ }\href@noop
  {} {\bibfield  {journal} {\bibinfo  {journal} {J. Alloy Compd.}\ }\textbf
  {\bibinfo {volume} {488}},\ \bibinfo {pages} {591} (\bibinfo {year}
  {2009})}\BibitemShut {NoStop}\bibitem [{\citenamefont {Hu}\ \emph {et~al.}(2011)\citenamefont {Hu},
  \citenamefont {Reid}, \citenamefont {Duan}, \citenamefont {Xia},\ and\
  \citenamefont {Yin}}]{Hu2011JPCM045501}  \BibitemOpen
  \bibfield  {author} {\bibinfo {author} {\bibfnamefont {L.}~\bibnamefont
  {Hu}}, \bibinfo {author} {\bibfnamefont {M.~F.}\ \bibnamefont {Reid}},
  \bibinfo {author} {\bibfnamefont {C.~K.}\ \bibnamefont {Duan}}, \bibinfo
  {author} {\bibfnamefont {S.}~\bibnamefont {Xia}}, \ and\ \bibinfo {author}
  {\bibfnamefont {M.}~\bibnamefont {Yin}},\ }\href@noop {} {\bibfield
  {journal} {\bibinfo  {journal} {J. Phys.: Condens. Matter}\ }\textbf
  {\bibinfo {volume} {23}},\ \bibinfo {pages} {045501} (\bibinfo {year}
  {2011})}\BibitemShut {NoStop}\bibitem [{\citenamefont {Gaigalas}\ \emph {et~al.}(2009)\citenamefont
  {Gaigalas}, \citenamefont {Gaidamauskas}, \citenamefont {Rudzikas},
  \citenamefont {Magnani},\ and\ \citenamefont
  {Caciuffo}}]{Gaigalas2009LJP403}  \BibitemOpen
  \bibfield  {author} {\bibinfo {author} {\bibfnamefont {G.}~\bibnamefont
  {Gaigalas}}, \bibinfo {author} {\bibfnamefont {E.}~\bibnamefont
  {Gaidamauskas}}, \bibinfo {author} {\bibfnamefont {Z.}~\bibnamefont
  {Rudzikas}}, \bibinfo {author} {\bibfnamefont {N.}~\bibnamefont {Magnani}}, \
  and\ \bibinfo {author} {\bibfnamefont {R.}~\bibnamefont {Caciuffo}},\
  }\href@noop {} {\bibfield  {journal} {\bibinfo  {journal} {Lithuanian J.
  Phys.}\ }\textbf {\bibinfo {volume} {49}},\ \bibinfo {pages} {403} (\bibinfo
  {year} {2009})}\BibitemShut {NoStop}\bibitem [{\citenamefont {Nov{\'a}k}\ and\ \citenamefont
  {Kuriplach}(1994)}]{Novak1994PRB2085}  \BibitemOpen
  \bibfield  {author} {\bibinfo {author} {\bibfnamefont {P.}~\bibnamefont
  {Nov{\'a}k}}\ and\ \bibinfo {author} {\bibfnamefont {J.}~\bibnamefont
  {Kuriplach}},\ }\href@noop {} {\bibfield  {journal} {\bibinfo  {journal}
  {Phys. Rev. B}\ }\textbf {\bibinfo {volume} {50}},\ \bibinfo {pages} {2085}
  (\bibinfo {year} {1994})}\BibitemShut {NoStop}\bibitem [{\citenamefont {Colarieti-Tosti}\ \emph {et~al.}(2002)\citenamefont
  {Colarieti-Tosti}, \citenamefont {Eriksson}, \citenamefont {Nordstrom},
  \citenamefont {Wills},\ and\ \citenamefont
  {Brooks}}]{Colarieti-Tosti2002PRB195102}  \BibitemOpen
  \bibfield  {author} {\bibinfo {author} {\bibfnamefont {M.}~\bibnamefont
  {Colarieti-Tosti}}, \bibinfo {author} {\bibfnamefont {O.}~\bibnamefont
  {Eriksson}}, \bibinfo {author} {\bibfnamefont {L.}~\bibnamefont {Nordstrom}},
  \bibinfo {author} {\bibfnamefont {J.}~\bibnamefont {Wills}}, \ and\ \bibinfo
  {author} {\bibfnamefont {M.~S.~S.}\ \bibnamefont {Brooks}},\ }\href@noop {}
  {\bibfield  {journal} {\bibinfo  {journal} {Phys. Rev. B}\ }\textbf {\bibinfo
  {volume} {65}},\ \bibinfo {pages} {195102} (\bibinfo {year}
  {2002})}\BibitemShut {NoStop}\bibitem [{\citenamefont {Zhou}\ and\ \citenamefont
  {Ozolins}(2011)}]{Zhou2011PRB085106}  \BibitemOpen
  \bibfield  {author} {\bibinfo {author} {\bibfnamefont {F.}~\bibnamefont
  {Zhou}}\ and\ \bibinfo {author} {\bibfnamefont {V.}~\bibnamefont {Ozolins}},\
  }\href@noop {} {\bibfield  {journal} {\bibinfo  {journal} {Phys. Rev. B}\
  }\textbf {\bibinfo {volume} {83}},\ \bibinfo {pages} {085106} (\bibinfo
  {year} {2011})}\BibitemShut {NoStop}\bibitem [{\citenamefont {Zhou}\ and\ \citenamefont
  {Ozolins}(2012)}]{Zhou2012PRB075124}  \BibitemOpen
  \bibfield  {author} {\bibinfo {author} {\bibfnamefont {F.}~\bibnamefont
  {Zhou}}\ and\ \bibinfo {author} {\bibfnamefont {V.}~\bibnamefont {Ozolins}},\
  }\href@noop {} {\bibfield  {journal} {\bibinfo  {journal} {Phys. Rev. B}\
  }\textbf {\bibinfo {volume} {85}},\ \bibinfo {pages} {075124} (\bibinfo
  {year} {2012})}\BibitemShut {NoStop}\bibitem [{\citenamefont {Anisimov}\ \emph {et~al.}(1991)\citenamefont
  {Anisimov}, \citenamefont {Zaanen},\ and\ \citenamefont
  {Andersen}}]{Anisimov1991PRB943}  \BibitemOpen
  \bibfield  {author} {\bibinfo {author} {\bibfnamefont {V.~I.}\ \bibnamefont
  {Anisimov}}, \bibinfo {author} {\bibfnamefont {J.}~\bibnamefont {Zaanen}}, \
  and\ \bibinfo {author} {\bibfnamefont {O.~K.}\ \bibnamefont {Andersen}},\
  }\href@noop {} {\bibfield  {journal} {\bibinfo  {journal} {Phys. Rev. B}\
  }\textbf {\bibinfo {volume} {44}},\ \bibinfo {pages} {943} (\bibinfo {year}
  {1991})}\BibitemShut {NoStop}\bibitem [{\citenamefont {Zhou}\ and\ \citenamefont
  {Ozolins}(2009)}]{Zhou2009PRB125127}  \BibitemOpen
  \bibfield  {author} {\bibinfo {author} {\bibfnamefont {F.}~\bibnamefont
  {Zhou}}\ and\ \bibinfo {author} {\bibfnamefont {V.}~\bibnamefont {Ozolins}},\
  }\href@noop {} {\bibfield  {journal} {\bibinfo  {journal} {Phys. Rev. B}\
  }\textbf {\bibinfo {volume} {80}},\ \bibinfo {pages} {125127} (\bibinfo
  {year} {2009})}\BibitemShut {NoStop}\bibitem [{\citenamefont {Nov{\'a}k}\ \emph
  {et~al.}(2013{\natexlab{a}})\citenamefont {Nov{\'a}k}, \citenamefont
  {Kn{\'\i}{\v z}ek},\ and\ \citenamefont {Kunes}}]{Novak2013PRB205139}  \BibitemOpen
  \bibfield  {author} {\bibinfo {author} {\bibfnamefont {P.}~\bibnamefont
  {Nov{\'a}k}}, \bibinfo {author} {\bibfnamefont {K.}~\bibnamefont {Kn{\'\i}{\v
  z}ek}}, \ and\ \bibinfo {author} {\bibfnamefont {J.}~\bibnamefont {Kunes}},\
  }\href@noop {} {\bibfield  {journal} {\bibinfo  {journal} {Phys. Rev. B}\
  }\textbf {\bibinfo {volume} {87}},\ \bibinfo {pages} {205139} (\bibinfo
  {year} {2013}{\natexlab{a}})}\BibitemShut {NoStop}\bibitem [{\citenamefont {Nov{\'a}k}\ \emph
  {et~al.}(2013{\natexlab{b}})\citenamefont {Nov{\'a}k}, \citenamefont
  {Kn{\'\i}{\v z}ek}, \citenamefont {Marysko}, \citenamefont {Jirak},\ and\
  \citenamefont {Kunes}}]{Novak2013JPCM446001}  \BibitemOpen
  \bibfield  {author} {\bibinfo {author} {\bibfnamefont {P.}~\bibnamefont
  {Nov{\'a}k}}, \bibinfo {author} {\bibfnamefont {K.}~\bibnamefont {Kn{\'\i}{\v
  z}ek}}, \bibinfo {author} {\bibfnamefont {M.}~\bibnamefont {Marysko}},
  \bibinfo {author} {\bibfnamefont {Z.}~\bibnamefont {Jirak}}, \ and\ \bibinfo
  {author} {\bibfnamefont {J.}~\bibnamefont {Kunes}},\ }\href@noop {}
  {\bibfield  {journal} {\bibinfo  {journal} {J. Phys.: Condens. Matter}\
  }\textbf {\bibinfo {volume} {25}},\ \bibinfo {pages} {446001} (\bibinfo
  {year} {2013}{\natexlab{b}})}\BibitemShut {NoStop}\bibitem [{\citenamefont {Nov{\'a}k}\ \emph
  {et~al.}(2014{\natexlab{a}})\citenamefont {Nov{\'a}k}, \citenamefont
  {Nekvasil},\ and\ \citenamefont {Kn{\'\i}{\v z}ek}}]{Novak2014JMMM228}  \BibitemOpen
  \bibfield  {author} {\bibinfo {author} {\bibfnamefont {P.}~\bibnamefont
  {Nov{\'a}k}}, \bibinfo {author} {\bibfnamefont {V.}~\bibnamefont {Nekvasil}},
  \ and\ \bibinfo {author} {\bibfnamefont {K.}~\bibnamefont {Kn{\'\i}{\v
  z}ek}},\ }\href@noop {} {\bibfield  {journal} {\bibinfo  {journal} {J. Magn.
  Magn. Mater.}\ }\textbf {\bibinfo {volume} {358-359}},\ \bibinfo {pages}
  {228} (\bibinfo {year} {2014}{\natexlab{a}})}\BibitemShut {NoStop}\bibitem [{\citenamefont {Nov{\'a}k}\ \emph
  {et~al.}(2014{\natexlab{b}})\citenamefont {Nov{\'a}k}, \citenamefont
  {Kunes},\ and\ \citenamefont {Kn{\'\i}{\v z}ek}}]{Novak2014OM414}  \BibitemOpen
  \bibfield  {author} {\bibinfo {author} {\bibfnamefont {P.}~\bibnamefont
  {Nov{\'a}k}}, \bibinfo {author} {\bibfnamefont {J.}~\bibnamefont {Kunes}}, \
  and\ \bibinfo {author} {\bibfnamefont {K.}~\bibnamefont {Kn{\'\i}{\v z}ek}},\
  }\href@noop {} {\bibfield  {journal} {\bibinfo  {journal} {Optical
  Materials}\ }\textbf {\bibinfo {volume} {37}},\ \bibinfo {pages} {414}
  (\bibinfo {year} {2014}{\natexlab{b}})}\BibitemShut {NoStop}\bibitem [{\citenamefont {Stevens}(1952)}]{Stevens1952PPSA209}  \BibitemOpen
  \bibfield  {author} {\bibinfo {author} {\bibfnamefont {K.~W.~H.}\
  \bibnamefont {Stevens}},\ }\href@noop {} {\bibfield  {journal} {\bibinfo
  {journal} {Proc. Phys. Soc. A}\ }\textbf {\bibinfo {volume} {65}},\ \bibinfo
  {pages} {209} (\bibinfo {year} {1952})}\BibitemShut {NoStop}\bibitem [{\citenamefont {Perdew}\ \emph {et~al.}(1996)\citenamefont {Perdew},
  \citenamefont {Burke},\ and\ \citenamefont {Ernzerhof}}]{Perdew1996PRL3865}  \BibitemOpen
  \bibfield  {author} {\bibinfo {author} {\bibfnamefont {J.~P.}\ \bibnamefont
  {Perdew}}, \bibinfo {author} {\bibfnamefont {K.}~\bibnamefont {Burke}}, \
  and\ \bibinfo {author} {\bibfnamefont {M.}~\bibnamefont {Ernzerhof}},\
  }\href@noop {} {\bibfield  {journal} {\bibinfo  {journal} {Phys. Rev. Lett.}\
  }\textbf {\bibinfo {volume} {77}},\ \bibinfo {pages} {3865} (\bibinfo {year}
  {1996})}\BibitemShut {NoStop}\bibitem [{\citenamefont {Blochl}(1994)}]{Blochl1994PRB17953}  \BibitemOpen
  \bibfield  {author} {\bibinfo {author} {\bibfnamefont {P.~E.}\ \bibnamefont
  {Blochl}},\ }\href@noop {} {\bibfield  {journal} {\bibinfo  {journal} {Phys.
  Rev. B}\ }\textbf {\bibinfo {volume} {50}},\ \bibinfo {pages} {17953}
  (\bibinfo {year} {1994})}\BibitemShut {NoStop}\bibitem [{\citenamefont {Kresse}\ and\ \citenamefont
  {Joubert}(1999)}]{Kresse1999PRB1758}  \BibitemOpen
  \bibfield  {author} {\bibinfo {author} {\bibfnamefont {G.}~\bibnamefont
  {Kresse}}\ and\ \bibinfo {author} {\bibfnamefont {D.}~\bibnamefont
  {Joubert}},\ }\href@noop {} {\bibfield  {journal} {\bibinfo  {journal} {Phys.
  Rev. B}\ }\textbf {\bibinfo {volume} {59}},\ \bibinfo {pages} {1758}
  (\bibinfo {year} {1999})}\BibitemShut {NoStop}\bibitem [{\citenamefont {Bryknar}\ \emph {et~al.}(2000)\citenamefont
  {Bryknar}, \citenamefont {Trepakov}, \citenamefont {Pot{\r u}{\v c}ek},\ and\
  \citenamefont {Jastrab{\'\i}k}}]{Bryknar2000JL605}  \BibitemOpen
  \bibfield  {author} {\bibinfo {author} {\bibfnamefont {Z.}~\bibnamefont
  {Bryknar}}, \bibinfo {author} {\bibfnamefont {V.}~\bibnamefont {Trepakov}},
  \bibinfo {author} {\bibfnamefont {Z.}~\bibnamefont {Pot{\r u}{\v c}ek}}, \
  and\ \bibinfo {author} {\bibfnamefont {L.}~\bibnamefont {Jastrab{\'\i}k}},\
  }\href@noop {} {\bibfield  {journal} {\bibinfo  {journal} {J. Lumin.}\
  }\textbf {\bibinfo {volume} {87-89}},\ \bibinfo {pages} {605} (\bibinfo
  {year} {2000})}\BibitemShut {NoStop}\bibitem [{\citenamefont {Brik}\ and\ \citenamefont
  {Srivastava}(2013{\natexlab{a}})}]{Brik2013EJSSSTR148}  \BibitemOpen
  \bibfield  {author} {\bibinfo {author} {\bibfnamefont {M.~G.}\ \bibnamefont
  {Brik}}\ and\ \bibinfo {author} {\bibfnamefont {A.~M.}\ \bibnamefont
  {Srivastava}},\ }\href@noop {} {\bibfield  {journal} {\bibinfo  {journal}
  {ECS Journal of Solid State Science and Technology}\ }\textbf {\bibinfo
  {volume} {2}},\ \bibinfo {pages} {R148} (\bibinfo {year}
  {2013}{\natexlab{a}})}\BibitemShut {NoStop}\bibitem [{\citenamefont {Brik}\ and\ \citenamefont
  {Srivastava}(2013{\natexlab{b}})}]{Brik2013JL69}  \BibitemOpen
  \bibfield  {author} {\bibinfo {author} {\bibfnamefont {M.~G.}\ \bibnamefont
  {Brik}}\ and\ \bibinfo {author} {\bibfnamefont {A.~M.}\ \bibnamefont
  {Srivastava}},\ }\href@noop {} {\bibfield  {journal} {\bibinfo  {journal} {J.
  Lumin.}\ }\textbf {\bibinfo {volume} {133}},\ \bibinfo {pages} {69} (\bibinfo
  {year} {2013}{\natexlab{b}})}\BibitemShut {NoStop}\bibitem [{\citenamefont {Xu}\ and\ \citenamefont
  {Adachi}(2009)}]{Xu2009JAP013525}  \BibitemOpen
  \bibfield  {author} {\bibinfo {author} {\bibfnamefont {Y.~K.}\ \bibnamefont
  {Xu}}\ and\ \bibinfo {author} {\bibfnamefont {S.}~\bibnamefont {Adachi}},\
  }\href@noop {} {\bibfield  {journal} {\bibinfo  {journal} {J. Appl. Phys.}\
  }\textbf {\bibinfo {volume} {105}},\ \bibinfo {pages} {013525} (\bibinfo
  {year} {2009})}\BibitemShut {NoStop}\bibitem [{\citenamefont {Curie}\ \emph {et~al.}(1974)\citenamefont {Curie},
  \citenamefont {Barthou},\ and\ \citenamefont {Canny}}]{Curie1974JCP3048}  \BibitemOpen
  \bibfield  {author} {\bibinfo {author} {\bibfnamefont {D.}~\bibnamefont
  {Curie}}, \bibinfo {author} {\bibfnamefont {C.}~\bibnamefont {Barthou}}, \
  and\ \bibinfo {author} {\bibfnamefont {B.}~\bibnamefont {Canny}},\
  }\href@noop {} {\bibfield  {journal} {\bibinfo  {journal} {J. Chem. Phys.}\
  }\textbf {\bibinfo {volume} {61}},\ \bibinfo {pages} {3048} (\bibinfo {year}
  {1974})}\BibitemShut {NoStop}\bibitem [{\citenamefont {Palumbo}\ and\ \citenamefont
  {Brown}(1970)}]{Palumbo1970JES1184}  \BibitemOpen
  \bibfield  {author} {\bibinfo {author} {\bibfnamefont {D.~T.}\ \bibnamefont
  {Palumbo}}\ and\ \bibinfo {author} {\bibfnamefont {J.~J.}\ \bibnamefont
  {Brown}},\ }\href@noop {} {\bibfield  {journal} {\bibinfo  {journal} {J.
  Electrochem. Soc.}\ }\textbf {\bibinfo {volume} {117}},\ \bibinfo {pages}
  {1184} (\bibinfo {year} {1970})}\BibitemShut {NoStop}\bibitem [{\citenamefont {Palumbo}\ and\ \citenamefont
  {Brown}(1971)}]{Palumbo1971JES1159}  \BibitemOpen
  \bibfield  {author} {\bibinfo {author} {\bibfnamefont {D.~T.}\ \bibnamefont
  {Palumbo}}\ and\ \bibinfo {author} {\bibfnamefont {J.~J.}\ \bibnamefont
  {Brown}},\ }\href@noop {} {\bibfield  {journal} {\bibinfo  {journal} {J.
  Electrochem. Soc.}\ }\textbf {\bibinfo {volume} {118}},\ \bibinfo {pages}
  {1159} (\bibinfo {year} {1971})}\BibitemShut {NoStop}\bibitem [{\citenamefont {Casta{\~n}eda}\ \emph {et~al.}(2005)\citenamefont
  {Casta{\~n}eda}, \citenamefont {Mu{\~n}oz~H},\ and\ \citenamefont
  {Caldi{\~n}o}}]{Castaneda2005OM1456}  \BibitemOpen
  \bibfield  {author} {\bibinfo {author} {\bibfnamefont {D.}~\bibnamefont
  {Casta{\~n}eda}}, \bibinfo {author} {\bibfnamefont {G.}~\bibnamefont
  {Mu{\~n}oz~H}}, \ and\ \bibinfo {author} {\bibfnamefont {U.}~\bibnamefont
  {Caldi{\~n}o}},\ }\href@noop {} {\bibfield  {journal} {\bibinfo  {journal}
  {Optical Materials}\ }\textbf {\bibinfo {volume} {27}},\ \bibinfo {pages}
  {1456} (\bibinfo {year} {2005})}\BibitemShut {NoStop}\bibitem [{\citenamefont {Whiffen}\ \emph {et~al.}(2014)\citenamefont
  {Whiffen}, \citenamefont {Jovanovi{\'c}}, \citenamefont {Anti{\'c}},
  \citenamefont {B{\'a}rtov{\'a}}, \citenamefont {Milivojevi{\'c}},
  \citenamefont {Drami{\'c}anin},\ and\ \citenamefont
  {Brik}}]{Whiffen2014JL133}  \BibitemOpen
  \bibfield  {author} {\bibinfo {author} {\bibfnamefont {R.~M.~K.}\
  \bibnamefont {Whiffen}}, \bibinfo {author} {\bibfnamefont {D.~J.}\
  \bibnamefont {Jovanovi{\'c}}}, \bibinfo {author} {\bibfnamefont {{\v
  Z}.}~\bibnamefont {Anti{\'c}}}, \bibinfo {author} {\bibfnamefont
  {B.}~\bibnamefont {B{\'a}rtov{\'a}}}, \bibinfo {author} {\bibfnamefont
  {D.}~\bibnamefont {Milivojevi{\'c}}}, \bibinfo {author} {\bibfnamefont
  {M.~D.}\ \bibnamefont {Drami{\'c}anin}}, \ and\ \bibinfo {author}
  {\bibfnamefont {M.~G.}\ \bibnamefont {Brik}},\ }\href@noop {} {\bibfield
  {journal} {\bibinfo  {journal} {J. Lumin.}\ }\textbf {\bibinfo {volume}
  {146}},\ \bibinfo {pages} {133} (\bibinfo {year} {2014})}\BibitemShut
  {NoStop}\bibitem [{\citenamefont {Chen}\ \emph {et~al.}(2001)\citenamefont {Chen},
  \citenamefont {Sammynaiken}, \citenamefont {Huang}, \citenamefont {Malm},
  \citenamefont {Wallenberg}, \citenamefont {Bovin}, \citenamefont {Zwiller},\
  and\ \citenamefont {Kotov}}]{Chen2001JAP1120}  \BibitemOpen
  \bibfield  {author} {\bibinfo {author} {\bibfnamefont {W.}~\bibnamefont
  {Chen}}, \bibinfo {author} {\bibfnamefont {R.}~\bibnamefont {Sammynaiken}},
  \bibinfo {author} {\bibfnamefont {Y.}~\bibnamefont {Huang}}, \bibinfo
  {author} {\bibfnamefont {J.-O.}\ \bibnamefont {Malm}}, \bibinfo {author}
  {\bibfnamefont {R.}~\bibnamefont {Wallenberg}}, \bibinfo {author}
  {\bibfnamefont {J.-O.}\ \bibnamefont {Bovin}}, \bibinfo {author}
  {\bibfnamefont {V.}~\bibnamefont {Zwiller}}, \ and\ \bibinfo {author}
  {\bibfnamefont {N.~A.}\ \bibnamefont {Kotov}},\ }\href@noop {} {\bibfield
  {journal} {\bibinfo  {journal} {J. Appl. Phys.}\ }\textbf {\bibinfo {volume}
  {89}},\ \bibinfo {pages} {1120} (\bibinfo {year} {2001})}\BibitemShut
  {NoStop}\end{thebibliography}
\end{document}